# Output coupling of a Bose-Einstein condensate formed in a TOP trap

J. L. Martin, C. R. McKenzie, N. R. Thomas, J. C. Sharpe, D. M. Warrington, P. J. Manson, W. J. Sandle and A. C. Wilson

*Department of Physics, University of Otago, PO Box 56, Dunedin, New Zealand*

Two distinct mechanisms are investigated for transferring a pure $^{87}$Rb Bose-Einstein condensate in the $|F = 2, m_F = 2\rangle$ state into a mixture of condensates in all the $m_F$ states within the $F = 2$ manifold. Some of these condensates remain trapped whilst others are output coupled in the form of an elementary pulsed atom laser. Here we present details of the condensate preparation and results of the two condensate output coupling schemes. The first scheme is a radio frequency technique which allows controllable transfer into available $m_F$ states, and the second makes use of Majorana spin flips to equally populate all the manifold sub-states.

## 1. Introduction

The production of Bose-Einstein condensation (BEC) with dilute alkali vapours [1] has generated a great deal of interest in coherent matter. Early experiments focussed on the properties of macroscopic atomic systems in a single quantum state. These included mean-field energy effects [1], careful analysis of the condensate transition point [2], condensate growth [3], propagation of sound waves [4], and collective excitations [5]. The measurement of dynamical features was improved by non-destructive probing schemes [6]. Realisation of an atom laser by output coupling from a Bose condensate [7] highlighted the importance of BEC, and the interference of two condensates [8] confirmed the presence of long range coherence. More recently, the production of multi-component condensates has revealed intriguing quantum fluid dynamics, and enabled precise measurement of relative quantum phase [9]. Experiments at JILA have focused on mixtures involving atoms in different ground state hyperfine levels (quantum number $F$), and coupling with a two photon (microwave plus radio frequency) transition. Experiments at MIT have used an optical dipole trap to confine a condensate occupying all magnetic sub-states (quantum number $m_F$) of the same hyperfine level [10]. Spin exchange processes result in domain formation which exhibits an anti-ferromagnetic interaction [11]. An important aspect of that work is the confinement of condensed atoms in sub-states which cannot be magnetically trapped.

In this paper we present the results of two techniques for transforming a single state $|F = 2, m_F = 2\rangle$ $^{87}$Rb Bose condensate into a mixture of all five magnetic sub-states of the $F = 2$ hyperfine level. Two of these states are magnetically confined ($|2, 2\rangle$ and $|2, 1\rangle$), with magnetic moments differing by a factor of 2, and the other states are unconfined.

Applying an RF field similar to that used in the evaporative cooling stage of the experiment we can couple atoms between adjacent $m_F$ states. It is possible to control the number of atoms which are coupled from the $|2, 2\rangle$ state into the other $m_F$ states: condensates with predetermined sub-state populations can be constructed. For example, we can limit the transfer into untrapped states.

In addition we have demonstrated the use of Majorana spin flips to couple atoms into all different $m_F$ states. This is achieved by removing the rotating bias field of a Time-averaged Orbiting Potential (TOP) trap [13], thereby placing the zero field point of a quadrupole trap at the centre of the condensate. This has the effect of rapidly spin flipping atoms into all available $m_F$ states, producing five equally populated condensates.

## 2. Making a Bose Condensate

The initial $|2, 2\rangle$ $^{87}$Rb Bose condensate is produced using laser cooling in a double magneto-optical trap (MOT) [14], and a combination of Majorana spin flip and radio frequency techniques for evaporative cooling [15] of atoms in a TOP trap. Our procedure is similar to that used elsewhere (for example [16]), but for critical comparison the experimental details are given below.

Light for two magneto-optical traps is derived from an external cavity diode laser and a semiconductor tapered amplifier, in a MOPA configuration [17]. The MOPA output power is set to 300 mW (although 500 mW is available), and the beam is split in two, one for each of the MOTs. Each beam is double passed through an acousto-optic modulator (AOM), giving precise computer control of the frequency and the intensity. The AOMs shift the frequency of each beam close to the cycling transition $F = 2 \rightarrow F' = 3$, and allow a detuning range of ±60 MHz.

The refrigeration process begins with a MOT formed in a region of relatively high rubidium vapour pressure ($10^{-8}$ Torr) to provide rapid collection of laser cooled atoms. Using the standard configuration of three orthogonal retro-reflected beams, in our case each with a diameter of 2 cm and a power of 10 mW, we load approximately $5 \times 10^8$ atoms in 0.5 s. These atoms are then transferred to a 'storage' MOT formed in a region of the vacuum system with a very low pressure suitable for magnetic trapping. The transfer process involves first switching off the cooling and trapping light and removing the weak magnetic field gradient, and then applying a 1.55 ms pulse of resonant laser light which 'pushes' the atoms along a 30 cm long, 15 mm diameter tube. To assist with the transfer efficiency, the resonant transfer pulse is circularly polarised to optically pump the atoms into the $|2, 2\rangle$ magnetically confined sub-state, and a confining hexapole magnetic guiding field is formed around the transfer tube using strips of flexible permanent magnet. The narrow transfer tube, an adjustable constriction and a differential pumping scheme provide an essentially rubidium free environment and background pressure below $10^{-11}$ Torr at the magnetic trap.

The storage MOT is formed in a small (70 × 20 × 20 mm) quartz cell pumped by a 55 L/s ion pump and a titanium sublimation pump. We estimate the transfer efficiency between MOTs to be approximately 20%. The push beam is not perfectly aligned along the axis of the transfer tube, but is focused to a waist just beside the storage MOT so as to avoid disrupting the trapped atoms. The collection and transfer process can be repeated many times to accumulate a large sample of atoms in the storage MOT, but is eventually limited by light assisted collisions. In our case, the storage MOT is formed using six independent beams (to reduce the effect of strong absorption of the cooling and trapping light by the large cloud), each with a power of 1.6 mW, a diameter of 1.5 cm and a detuning of –16 MHz, giving a MOT lifetime of 70 s. This allows us to transfer up to 50 atomic clouds without significant loss in efficiency, giving approximately $5 \times 10^9$ atoms in the storage MOT.

Once the multiple transfer process is complete the atoms are compressed and cooled in optical molasses [18] to maximise the phase space density in the storage MOT, and then optically pumped into the |2, 2⟩ state before being loaded into a magnetic trap for evaporative cooling. Compression in the storage MOT involves increasing the axial magnetic field gradient from 17 G cm$^{-1}$ to 30 G cm$^{-1}$ for 20 ms. Cooling in molasses takes an additional 20 ms, and involves reducing the axial magnetic field gradient to 6 G cm$^{-1}$ and increasing the laser detuning to –26 MHz. Immediately after cooling in optical molasses, but before the magnetic trap is switched on, the atoms are optically pumped into the |2, 2⟩ state with a 30 µs circularly polarised resonant laser pulse and a weak bias field (formed using coils associated with the magnetic trap). The optical pumping gives a threefold increase in the number of magnetically trapped atoms.

We load the atoms into a time-averaged orbiting potential (TOP) trap [13], which consists of a quadrupole field and a rotating bias field (the TOP field). The quadrupole field is created by two water cooled coils each with 560 turns, producing a radial magnetic field gradient ($B'_q$) of up to 200 G cm$^{-1}$ at 10 A. The TOP field ($B_{TOP}$) is created by two pairs of Helmholtz coils each with 6 turns, driven in quadrature by a 1.2 kW stereo amplifier and transformers, providing up to 50 G rotating at 7 kHz. The TOP field eliminates trap loss from Majorana spin flips into untrapped magnetic sub-states, which occur at the zero field point of the quadrupole field, by displacing that point into an orbit beyond the extent of the cloud. This orbit is commonly known as the 'circle of death', since atoms along its path are spin flipped out of the trap, and it transpires that this mechanism can be used to perform efficient evaporative cooling. To confine atoms, the frequency of field rotation must be greater than the centre of mass oscillation frequency of the atoms (typically 10 to 100 Hz radially) thereby maintaining the peripheral orbit, but lower than the Larmor frequency (2.8 MHz in 4 G) so that the magnetic moments of the atoms remain aligned with the field. The spatial distribution of the atoms in the TOP trap conforms to the time-average of the rotating potential, which is harmonic with an axial frequency √8 times that in the radial direction. However, for RF coupling (and evaporation) one must consider the instantaneous magnetic field, *not* the time-averaged potential.

To load the magnetic trap the magnetic potential is mode-matched to the spatial distribution of the laser cooled cloud so that the phase space density is maintained and any subsequent oscillations minimised. In addition the radius of the 'circle of death' must exceed our typical cloud diameter of 4 mm. These conditions are fulfilled by using $B'_q$ = 75 G cm$^{-1}$ and $B_{TOP}$ = 40 G. Approximately 50% of the optically cooled atoms are confined by the magnetic field. We obtain approximately 10$^9$ atoms in the TOP trap with an average density of 3×10$^{10}$ atoms cm$^{-3}$ and a temperature of 150 µK, giving a phase space density of 3×10$^{-8}$ with a lifetime (limited by collisions with background atoms) in excess of 90 s. At this stage the elastic collision rate, which determines the rethermalisation rate [19], is estimated to be 3 s$^{-1}$. Next we evaporatively cool the atoms to achieve Bose condensation.

Our evaporative cooling sequence is a three step process:

i.  A linear increase in the radial quadrupole field $B'_q$ from 75 G cm$^{-1}$ to 200 G cm$^{-1}$ over a period of 16 s, changing the trap oscillation frequency from 9 Hz to 29 Hz.

ii. An exponential decrease of $B_{TOP}$ from 40 G to 4 G over a period of 24 s, changing the trap oscillation frequency from 29 Hz to 90 Hz.

iii. An exponential scan of an RF field from 5.6 MHz to 3.4 MHz over a period of 17 s, with an amplitude of 40 mG.

Note: all trap oscillation frequencies quoted are radial. Axial frequencies are √8 larger.

For large thermal clouds, we find that evaporating with the 'circle of death' is simple to implement and efficient because it simultaneously evaporates and increases the trap oscillation frequency (and therefore the elastic collision rate). However, closer to the BEC transition temperature, radio frequency evaporation gives more precise control. The radio frequency source is a GPIB controlled function generator, and the RF field is applied using a pair of single loop Helmholtz coils, oriented perpendicular to the plane of the rotating bias field, and driven by a 3 W amplifier.

Following evaporation we probe the atomic cloud by first increasing $B_{TOP}$ to 30 G in 10 μs to weaken the confining potential, before switching the quadrupole component of the magnetic trap off and performing a time-of-flight (TOF) measurement. This gives us a fast effective field decay time compared to the trap oscillation frequency. We probe with a 2 mW, 50 μs pulse of light resonant with the cycling transition, synchronised to the rotating bias field, and image the shadow formed in the probe beam with a 16-bit cooled CCD array camera. The number of atoms is calculated from the optical depth.

BEC is evident in absorptive images of the atomic cloud following release from the magnetic trap and time-of-flight expansion. Performing a series of evaporative cooling ramps with successively lower final RF frequencies, we eventually observe a sudden increase in the optical depth in the centre of the cloud (see Figure 1). Also, the time-of-flight images of clouds cooled below this transition point show elliptical velocity distributions with long axes aligned with the tightly confining axis of the magnetic trap. For the same time-of-flight conditions, clouds cooled just above the transition temperature show an isotropic velocity distribution. The transition temperature is approximately 400 nK, and at 200 nK we obtain approximately $2\times10^5$ atoms in a Bose condensate with no evidence of a thermal cloud.

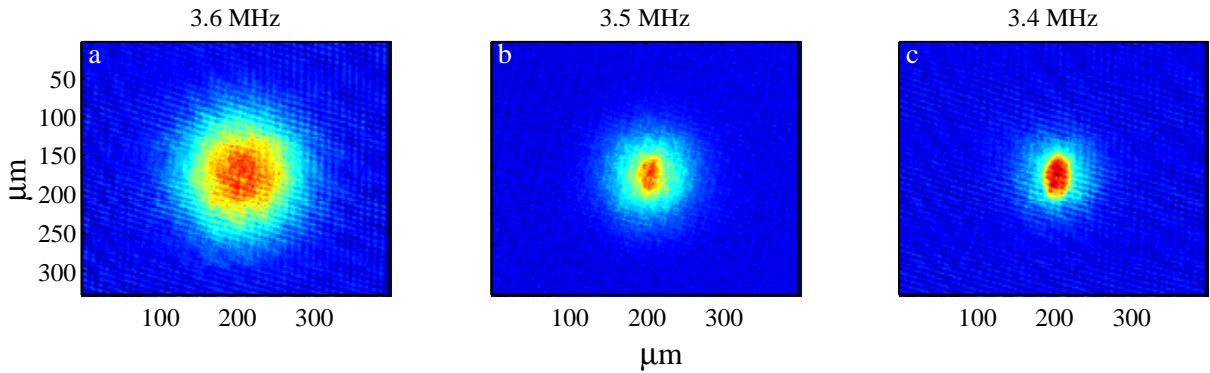

Figure 1    Time-of-flight images showing (a) a thermal cloud of atoms, (b) a mixed cloud of thermal and condensate atoms and (c) a condensate cloud. The final RF frequency for each evaporative cooling ramp is given above the corresponding image.

# 3. Output Coupling the Condensate

## 3.1 Radio Frequency Output Coupling

The first output coupling scheme we investigated with a Bose condensate in a TOP trap was to apply an RF pulse resonant with the atoms at the centre of the magnetic trap. This RF field transfers population between adjacent $m_F$ states, i.e. $|2, 2\rangle \rightarrow |2, 1\rangle \rightarrow |2, 0\rangle$ etc. This process is similar to that used by the MIT BEC group in their atom laser experiments [7]. However, in their system there are only three $m_F$ states in the $F = 1$ manifold. In our case there are five $m_F$ states and the dynamics of the coupling process are different. In particular there are two trapped states ($|2, 2\rangle$ and $|2, 1\rangle$) which may interact via mean field effects during and after application of the coupling field. These interactions may lead to phase shifts or decoherence between the two condensates.

The RF pulse is created in a manner similar to the evaporative cooling ramp. The RF coils create a magnetic field linearly polarised perpendicular to the plane of the rotating bias field. For transitions between $m_F$ states this field can be expressed as a sum of two circularly polarised fields. Each of these fields drives a transition with a Rabi frequency proportional to half the amplitude of the field created by the RF coils. RF magnetic field strengths quoted in this paper relate directly to $B_{rf}$ and not to the total magnetic field strength.

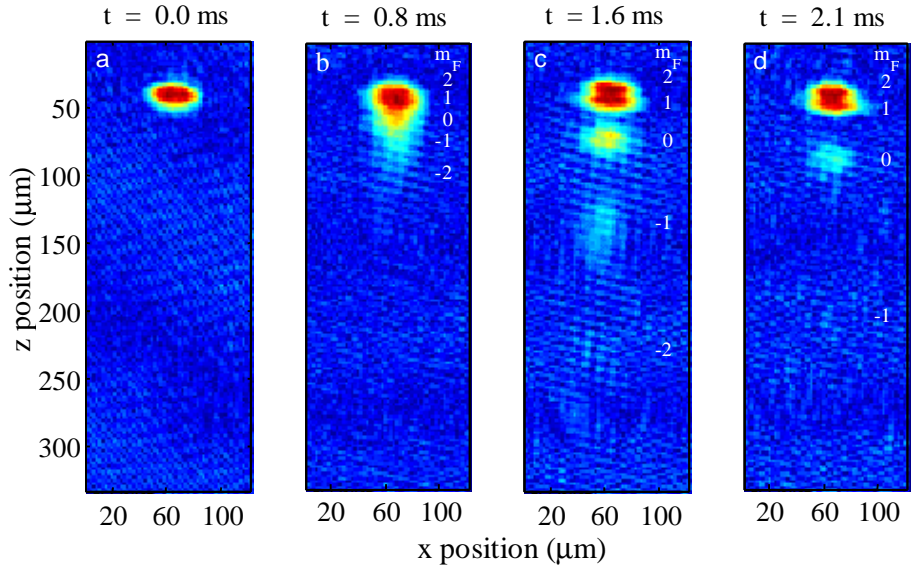

Figure 2   *In situ* images of the trap for a sequence of times after the RF pulse of 286µs has been applied. The $m_F$ labels indicate the position of the individual atomic states.

The length, timing and power of the RF pulse are all computer controlled, allowing fine adjustment of the coupling. For our final TOP field of $B_{TOP} = 4$ G, an RF field tuned to the Larmor frequency of 2.8 MHz couples atoms out of the original condensate. Because of a bandwidth limit on the oscillator's external control circuitry we investigated only pulses that are long compared to the TOP field rotation time. With such a long pulse one might expect spatially selective coupling from the centre of the trap. However, in a TOP trap the instantaneous coupling surface is one of constant quadrupolar magnetic field magnitude. This surface rotates and therefor couples atoms from all parts of the cloud.

Figure 2 shows the results of an output coupling pulse of 286 µs (equal to two TOP field rotations) and a field strength $B_{rf} = 50$ mG. This coupling pulse is strong enough to sequentially couple to all available $m_F$ states, but not strong enough to populate them equally. Because $g_F = +½$ in this case, atoms in states $|2, 1\rangle$ and $|2, 2\rangle$ are both trapped while $|2, 0\rangle$ atoms fall under gravity, and $|2, -1\rangle$ and $|2, -2\rangle$ atoms are repelled from the trap. Note that atoms in a negative $m_F$ state will be repelled downward. This is because atoms are initially displaced below the magnetic potential minimum due to gravity, and when the $m_F$ state changes that initial displacement leads to a downward force. The separation between the $|2, 1\rangle$ and $|2, 2\rangle$ clouds (with $|2, 1\rangle$ being lower) is due to their magnetic moments giving rise to different gravitational displacements.

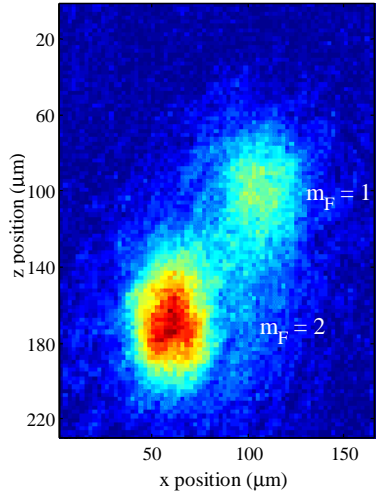

Figure 3   Time of flight image of the $|2, 1\rangle$ and $|2, 2\rangle$ components of the condensate. The elliptical shape of each component in TOF is a result of self mean-field repulsion.

If the atoms coupled into the two trapped states are allowed to expand ballistically then we expect to see two slightly separated components that have the same characteristic elliptical shape of the original condensate (such as in Figure 1c). Figure 3 shows such a time of flight image. The separation (both axial and horizontal) between the two components is due to the difference in their position with respect to the quadrupole field when the trap is relaxed, leading to different motion in the decaying quadrupole field. The important feature of this time of flight image is the ellipticity of both of the clouds, which is a result of self mean-field repulsion in the individual condensate components, confirming that the coupling produces a Bose condensate in the $|2, 1\rangle$ state while not destroying the $|2, 2\rangle$ 'parent' condensate. Since there are no appreciable thermal clouds formed, we know that there is no significant heating associated with the coupling process.

In order to observe the behaviour of the two trapped components we reduce the trapping frequency from 90 Hz to 66 Hz in 20 µs, setting the two components into centre of mass oscillation. Since the two components have different magnetic moments they experience different trapping potentials and hence have different oscillation frequencies. Figure 4 shows the position of the two components as a function of time. The position of each component is fitted with a variable phase sine wave and the straight dotted lines indicate the fitted equilibrium positions of the two clouds.

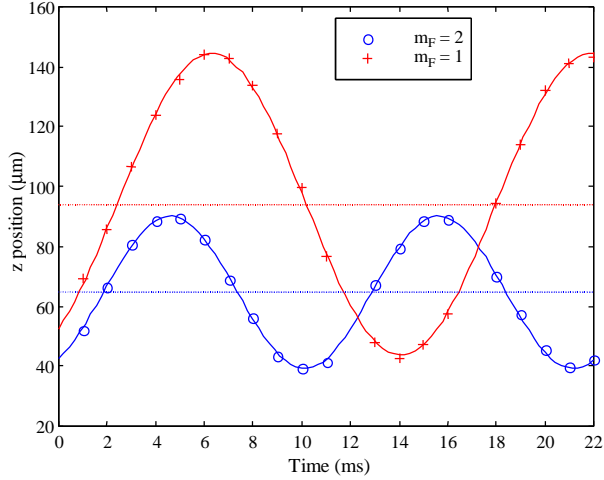

Figure 4  Centre of mass oscillation of the two states in the magnetic trap showing the effects of the different magnetic moments and gravity.

The fit to the data is extremely good. We expect the ratio of the two oscillation frequencies to be √2 since the magnetic moments (and hence the force) differ by a factor of 2. The experimental ratio is 1.40(3). Also the equilibrium separation of the two components and their particular oscillation frequencies can be calculated [13].

$$F_z = \frac{\partial U}{\partial z} = \frac{\partial}{\partial z}\left(2g_F m_F m_B Q z^2\right) = w^2 z \quad (1)$$

where $w^2 = 4g_F m_F m_B Q$ and $Q = \frac{B_q'^2}{B_{TOP}}$. To calculate the gravitational displacement of the atomic clouds from the magnetic potential minimum we equate $F_z$ to $mg$ (i.e. the z axis is parallel to gravity) giving

$$z_{disp} = \frac{mg}{w^2}$$

Evaluating this for $|2, 2\rangle$ gives $z_{disp} = 29.7$ μm and $w = 93$ Hz, and for $|2, 1\rangle$, $z_{disp} = 59.4$ μm and $w = 66$ Hz. The experimentally measured frequencies are 91(2) and 65(1) Hz respectively and the difference between the equilibrium positions is 29(1) μm. These results confirm that the two states are $|2, 1\rangle$ and $|2, 2\rangle$.

Finally, we do not observe any interaction between the two components when they collide (at times ~13 ms and ~17 ms in Figure 4). We intend to consider this in more detail in future work.

### 3.2  Zero Magnetic Field Output Coupling

The second output coupling scheme we have considered uses Majorana spin flips between $m_F$ states in the $F = 2$ manifold to couple atoms out of the original Bose condensate state. Majorana spin flips occur in regions of the trap where the magnetic moment of the atom can not adiabatically follow changes in the magnetic field direction [20].

$$\frac{d\mathbf{q}_B}{dt} > \frac{\mathbf{m}_B B}{\hbar} = \mathbf{w}_{LARMOR} \qquad (2)$$

where $\mathbf{q}_B$ is the angle between the magnetic field and *z*-axis. When an atom enters the non-adiabatic region, the precession of its magnetic moment is much slower than the angular changes in the magnetic field. Hence the magnetic moment of an atom upon exiting the region will be approximately the same as when it entered. The final state of an atom is thus determined by projecting the initial magnetic moment of the atom (determined by the initial local magnetic field) onto the new quantisation axis defined by the final local magnetic field. Therefore, the final state will be uniquely determined by the atomic trajectory through the non-adiabatic region. To determine the final population distribution amongst the $m_F$ states we must integrate over all possible trajectories through the non-adiabatic region. Since any trajectory is equally probable and the magnetic field is quadrupolar each available $m_F$ state will be equally populated.

In a TOP trap the non-adiabatic region occurs where the Larmor frequency approaches zero i.e. centred about the zero field point in the magnetic field. Usually this is well beyond the cloud of atoms at the 'circle of death'. However if $B_{TOP}$ is reduced to zero then the field zero moves to the centre of the condensate and coupling to other states can occur.

Figure 5 shows the result of removing the TOP field. The transverse separation of the atoms is due to the *fast* removal of the TOP field and the *slow* turn-off of the quadrupole field. Removing the TOP field 'jumps' the quadrupole field into the centre of the trap giving the atoms a velocity in the direction of this motion. The velocity 'kick' is exactly the same for all atoms. The quadrupole field remains on for ~500 µs after the TOP field is removed. The quadrupole field is then turned off, but the field takes approximately 1.5 ms to decay due to the inductance of the coils, and during that time the different components are spatially separated by the Stern-Gerlach effect, as seen in Figure 5.

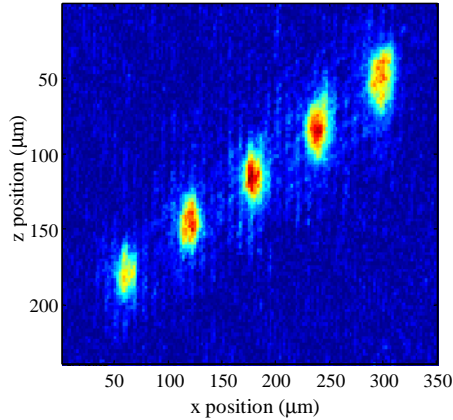

Figure 5    A 2 ms time of flight image of a condensate that has been coupled into all five available $m_F$ states by Majorana spin flips.

Note that each of the individual components in Figure 5 has approximately the same ellipticity. This shows that the self mean-field repulsion of each of the states is approximately the same and that any heating introduced by the coupling is significantly smaller than the mean-field energy of the condensates. A key feature of this result is that all the states are equally populated, as we predict.

From (2) we can derive a radius for the region in which Majorana spin flips may occur. The size of this coupling region is different at the centre of a quadrupole trap than at the 'circle of death' of a TOP trap. For a quadrupole trap the radius of the region, $r_{quad}$ is given by

$$r_{quad} \sim \sqrt{\frac{\hbar v}{m_B B'_q}}$$

where $v$ is the velocity of atoms traversing the zero field region. Using this expression the size of the non-adiabatic region in a TOP trap can be calculated by substituting the velocity of the zero field point for the velocity of the atoms, since atoms at the very edge of the cloud are nearly stationary. Thus for a TOP trap the coupling region, $r_{TOP}$, is given by

$$r_{TOP} \sim \sqrt{\frac{\hbar w_{TOP}}{m_B Q}}$$

where $Q$ is defined in (1) and $w_{TOP}$ is the TOP rotation frequency. For our TOP trap parameters ($w_{TOP} = 2\pi \times 7$ kHz, $B'_q = 75$ to 200 G cm$^{-1}$ and $B_{TOP} = 40$ to 4 G) $r_{TOP}$ varies between 33 µm and 4 µm during the evaporative cooling sequence. For output coupling in a quadrupole field with atoms at 400 nK, $r_{quad}$ is only ~200 nm. One might expect this to suggest that the coupling process would be slow since atoms would be unlikely to intersect such a small region. However, for $B'_q = 200$ G cm$^{-1}$ the change in magnetic field across this 'sink' is only 4 mG, so any residual AC fields larger than this will increase the effective size. We estimate that the minimum residual TOP field we can currently achieve is 20 mG, which increases the size of the quadrupole 'sink' to around 1 µm. Even with this larger size one might still expect spatially dependent coupling to occur for interaction times shorter than the trap oscillation period (1/90 Hz).

However, unlike an harmonic oscillator the restoring, force in a quadrupole trap is the same for all atoms so that the time required to reach the centre of the trap is different. This time is dependent on the absolute energy of an atom. If the condensate cloud has a radius $r$ then the maximum possible time for any atom to reach the centre of the trap is

$$t = \sqrt{\frac{8mr}{g_F m_F m_B B'_q}}$$

where $m$ is the mass of a $^{87}$Rb atom. For $B'_q = 200$ G cm$^{-1}$ and $r = 15$ µm (typical for our condensates) this time is 970 µs. (Note that this is five time faster than the equivalent time in our TOP trap). For our measurements the atoms remain in a quadrupole field for ~500 µs, followed by 1.5 ms decay of the quadrupole field, so that all of the atoms have time to interact with the zero field region. Thus all of the atoms have an opportunity to undergo a Majorana spin flip and couple into other states. Spatially selective output coupling with this technique is possible, but only with significantly faster control of the quadrupole field.

## 4. Conclusions

We have presented results of two methods for output coupling and producing multi-component Bose condensates in a TOP trap, and given a detailed description of the process used to form a $|2, 2\rangle$ Bose condensate.

We have shown that RF output coupling provides a degree of control over the relative condensate populations of each of the five available states. We used a centre of mass oscillation technique to look for interaction between the confined states ($|2, 2\rangle$, $|2, 1\rangle$) but did not observe any significant effects. We intend to perform a more detailed study of this.

We have given a detailed description and presented results of Majorana spin flip effects on a condensate in a quadrupole trap. We obtained five equally populated Bose condensates in each of the five possible $m_F$ states. From a time of flight image of these states it appears that the mean-field repulsion of all the states is approximately the same.

## Acknowledgements

We would like to thank Dr C. Foot at the University of Oxford, Assoc. Prof. R. Ballagh, Prof. C. Gardiner and the BEC Theory group at the University of Otago for their helpful advice. We are grateful for the financial support of the Royal Society of New Zealand Marsden Fund (contract: UOO508), the Foundation for Research, Science and Technology Postdoctoral Fellowship programme (contract: UOO524), Lottery Science, the Royal Society (London) and the University of Otago Research Committee.